\pdfoutput=1
\documentclass[12pt]{article}

\usepackage{amsmath,graphicx}
\usepackage{simplewick}
\usepackage{epsf}
\usepackage{graphicx,epsfig}
\usepackage{amsfonts}
\usepackage{amssymb}
\usepackage[nosort]{cite}
\usepackage{setspace}
\usepackage{float}

\def\CL{{\cal L}}
\def\CO{{\cal O}}

\def\tR{{\tilde R}}

\def\high{\vphantom{\Biggl(}\displaystyle}

\def\mpl{M_{\rm P}}

\makeatletter
\renewcommand\section{\@startsection {section}{1}{\z@}%
                                 {-3.5ex \@plus -1ex \@minus -.2ex}%
                                   {2.3ex \@plus.2ex}%
                                   {\normalfont\large\bfseries}}
\renewcommand\subsection{\@startsection{subsection}{2}{\z@}%
                                   {-3.25ex\@plus -1ex \@minus -.2ex}%
                                     {1.5ex \@plus .2ex}%
                                     {\normalfont\bfseries}}
\renewcommand\subsubsection{\@startsection{subsubsection}{3}{\z@}%
                                   {-3.25ex\@plus -1ex \@minus -.2ex}%
                                     {1.5ex \@plus .2ex}%
                                     {\normalfont\itshape}}
\makeatother

\newcommand{\Letter}{
\setlength{\textwidth}{16.5cm}
   \setlength{\textheight}{22.6cm}
    \hoffset=-0.5in
\voffset=-2.1cm }

\Letter

\begin{document}
\newcommand{\be}{\begin{equation}}
\newcommand{\ee}{\end{equation}}
\newcommand{\bea}{\begin{eqnarray}}
\newcommand{\eea}{\end{eqnarray}}
\newcommand{\barr}{\begin{array}}
\newcommand{\earr}{\end{array}}
\def\bal#1\eal{\begin{align}#1\end{align}}

\thispagestyle{empty}
\begin{flushright}
\end{flushright}

\vspace*{0.3in}
\begin{spacing}{1.1}

\begin{center}
{\large \bf Standard Clock in Primordial Density Perturbations
\\[0.05in]
and Cosmic Microwave Background}

\vspace*{0.3in} {Xingang Chen$^1$ and Mohammad Hossein Namjoo$^2$}
\\[.3in]
{\em
$^1$Department of Physics, The University of Texas at Dallas, Richardson, TX 75083, USA \\
$^2$School of Astronomy, Institute for Research in
Fundamental Sciences (IPM),\\
P.~O.~Box 19395-5531,
Tehran, Iran } \\[0.3in]
\end{center}

\begin{center}
{\bf
Abstract}
\end{center}
\noindent
Standard Clocks in the primordial epoch leave a special type of features in the primordial perturbations, which can be used to directly measure the scale factor of the primordial universe as a function of time $a(t)$, thus discriminating between inflation and alternatives. We have started to search for such signals in the Planck 2013 data using the key predictions of the Standard Clock. In this Letter, we summarize the key predictions of the Standard Clock and present an interesting candidate example in Planck 2013 data. Motivated by this candidate, we construct and compute full Standard Clock models and use the more complete prediction to make more extensive comparison with data. Although this candidate is not yet statistically significant, we use it to illustrate how Standard Clocks appear in Cosmic Microwave Background (CMB) and how they can be further tested by future data. We also use it to motivate more detailed theoretical model building.

\vfill

\newpage
\setcounter{page}{1}


\newpage

\section{Introduction}
\setcounter{equation}{0}

Our understanding on the origin of the Universe has advanced considerably in recent years through interactions between experiments and theories. We have a large number and variety of ongoing and upcoming experiments that are mapping the entire observable universe. One of the most important achievements of these experiments is to produce, one way or the other, different and complimentary maps of the distributions of large scale structures, including various spectra and objects, in our Universe.
These maps are the gold mines to advance our knowledge in cosmology. All these large scale structures originated from some tiny fluctuations in the very early universe, the primordial perturbations. One of the most beautiful ideas in modern cosmology is that these perturbations are seeded by quantum fluctuations of fields present in an early epoch responsible for the Big Bang. By studying properties of these maps, we learn properties of this epoch, as well as fundamental physics in conditions that are inaccessible for experiments on Earth.

In the past two decades, the data from CMB and Large Scale Structures (LSS) strongly support the inflationary paradigm \cite{Guth:1980zm,Linde:1981mu,Albrecht:1982wi,Starobinsky:1980te,Sato:1980yn} as the leading candidate for this primordial epoch.
The simplest inflationary models predict the primordial perturbations to be superhorizon, approximately scale-invariant, adiabatic and Gaussian \cite{Mukhanov:1981xt,Hawking:1982cz,Starobinsky:1982ee,Guth:1982ec,Bardeen:1983qw}.
All of these have been verified to some extent by the results from the Wilkinson Microwave Anisotropy Probe (WMAP) \cite{Komatsu:2010fb} and the Planck satellite \cite{Ade:2013zuv,Ade:2013uln}. The properties of these perturbations are summarized quantitatively by two of the six parameters in the Standard Model of Cosmology, the $\Lambda$CDM model.

On the other hand, other possibilities have also been speculated as alternative theories to the inflationary scenario. From the perspective of theoretical model building, none of them has been as successful as inflation. See Ref.~\cite{Brandenberger:2012zb,Ijjas:2013vea,Guth:2013sya,Linde:2014nna} for the current status. Nonetheless, models may be improved or become complicated to fit the data. This is possible because there are only two parameters in the Standard Model that are relevant to the primordial epoch, leaving rooms for theoretical freedoms. Therefore an equally important approach in cosmology is to search for beyond-Standard-Model signals in data that can be used to distinguish different scenarios.

Phenomenologically one can distinguish four different kinds of primordial epochs, classified by the time dependence of the scale factor $a(t)\sim t^p$: the fast-expanding or fast-contracting scenarios, and the slowly-expanding or slowly-contracting scenarios. (The contracting scenarios require a bounce to match the Big Bang.) Each of them has a different fingerprint index in terms of the parameter $p$ \cite{Chen:2011zf,Chen:2011tu}. The acceleratedly-expanding scenario, namely inflation, has $|p|>1$; the fast-contracting scenario \cite{Wands:1998yp,Finelli:2001sr} has $p\sim \CO(1)<1$; the slowly-expanding scenario has $-1 \ll p <0$; and the slowly-contracting scenario \cite{Khoury:2001wf} has $0<p\ll 1$.\footnote{The case $p\sim \CO(-1) > -1$ ($-\infty<t<0$) is also acceleratedly expanding.}
For $p>1$, $t$ runs from $0$ to $+\infty$; for all other $p$, $t$ runs from $-\infty$ to $0$.
The choices of $t$ are based on the requirement that the quantum fluctuations in this epoch exit the horizon, so that they can give rise to the acoustic oscillations in the CMB after reentry during the Big Bang.

The primordial perturbations, which are seeded by quantum fluctuations in these epochs, consist of scalar and tensor modes. While the scalar mode determines the density perturbations at the beginning of the Big Bang as the source of the large scale structures, the tensor mode corresponds to the gravitational quantum fluctuations and records the magnitude of the Hubble parameter during the epoch. Therefore the tensor mode serves as a good discriminator between the scenarios with fast-evolving scale factor and those with slowly-evolving scale factor. In particular, if the tensor mode origin of the recent CMB B-mode detection by the BICEP2 experiment \cite{Ade:2014xna} is confirmed, both the slowly-expanding and slowly-contracting scenarios will be ruled out. Nonetheless, phenomenologically, the tensor mode does not distinguish the inflation from the fast-contracting scenarios. For example, both the inflation and the matter contraction can give rise to observable tensor mode with approximately scale-invariant spectra \cite{Wands:1998yp,Finelli:2001sr}.

In this letter, we consider a different type of observables. A main reason the degeneracy of scenarios could exist is that the observables we mentioned so far (namely the approximately scale-invariant scalar and tensor modes) are all convoluted consequences of the scale factor $a(t)$, the defining property of different scenarios. A direct measurement of $a(t)$ would provide an independent and direct evidence for a scenario, as was done for the late-time accelerated universe using the Standard Candles \cite{Perlmutter:1997zf,Riess:1998cb}. This turns out to be possible: oscillating massive fields in the primordial epoch can serve for this purpose as the Standard Clocks \cite{Chen:2011zf,Chen:2011tu,Chen:2012ja}.
The massive field oscillates with a frequency that can be thought of as ticks of a clock.
This Standard Clock imprints its ticks as a special type of features in the primordial perturbations, thereby letting some imprints in the CMB angular power spectra, the non-Gaussianities and the distribution of large scale structures.
The patterns of these ticks are a direct record of $a(t)$ of the primordial universe.

In this Letter, after summarising the main results of the theoretical proposal of the Standard Clock, we compare its key predictions with the Planck 2013 residual data. A full-scale comparison will be the subject of the next paper \cite{ToAppear}. Here we focus on one interesting candidate emerging from this comparison, although it is still not statistically significant. Motivated by this candidate we construct explicit Standard Clock models and compute the full power spectrum. This is a completion of the above key predictions, under the same number of model parameters. We again see encouraging signs after this prediction is compared with the Planck data.

\section{Standard Clocks}
\label{Sec:StandardClocks}

We start with a summary of the key requirements and properties of the Standard Clock \cite{Chen:2011zf,Chen:2011tu,Chen:2012ja}.
There are two requirements to have a Standard Clock in a primordial scenario:

\begin{enumerate}
  \item We need an extra massive field with mass much larger than the event-horizon mass-scale of the corresponding primordial epoch.  For example, for inflation this mass scale is the Hubble parameter. This massive field is excited classically by some sharp features and oscillating.
  \item This massive field starts to oscillate at least several efolds after the beginning of the {\em observable} primordial epoch. For example, for inflationary scenario, the observable scales include approximately 60-efolds towards the end of inflation; the massive field has to {\em start} oscillating at least a few efolds within the 60-efolds, but not at or before the beginning of these 60-efolds.
\end{enumerate}

Standard Clocks generate two qualitatively different types of signals in the primordial perturbations, which are connected to each other and contain different properties. It is important to classify them and sort out which properties can be most robustly used to measure $a(t)$, which are less robust but can be auxiliary, and which cannot \cite{Chen:2011zf,Chen:2011tu}.

The first type of signal is generated by the sharp feature that excites the massive field. Like all sharp feature signals, this signal has a characteristic sinusoidal running as a function of scales,
\bea
\sim \cos(K/k_0+{\rm phase}) ~,
\label{sin_part}
\eea
where $k_0$ is both the $K$-location of the sharp feature and the wavelength of the oscillation.
Here $K\equiv k_1+k_2=2k_1$ for power spectrum, $K\equiv k_1+k_2+k_3$ for bispectrum and so on.
In inflation models, examples of various sharp features have been studied in e.g.~Ref.~\cite{Starobinsky:1992ts,Adams:2001vc,Chen:2006xjb,Achucarro:2010da,Miranda:2012rm,Bartolo:2013exa}. But we emphasize that the statement here is stronger -- this leading order behavior also applies to non-inflationary scenarios \cite{Chen:2011zf,Chen:2011tu}.
The sinusoidal running itself cannot be used to measure $a(t)$ because it is qualitatively the same for all scenarios. This can be intuitively understood -- the sharp feature has only one click and it does not contain any clock information.

The second type of signal is generated by the oscillation of massive field after it is excited. The most important property of this signal is its characteristic resonant running, which we shall explain more with explicit examples shortly. The pattern between successive oscillations is determined by the ticks generated by the Standard Clock and the scale factor $a(t)$, which is unique for each scenario.

Overall, in the {\em full} Standard Clock signals, the following are the two most robust properties that can be used to distinguish different scenarios.
\begin{enumerate}
\item[A.] {\em The clock signal.} The fingerprint resonant running signal generated by the Standard Clock, as a fractional correction to the leading order approximately scale-invariant power spectrum, $\Delta P_\zeta/P_{\zeta 0}$, or as the leading order non-Gaussianities, is given by
    \bea
    C \left(\frac{K}{k_r} \right)^\alpha
    \sin \left[ \frac{p^2}{1-p} \omega \left( \frac{K}{k_r} \right)^{1/p} + \varphi \right] ~.
    \label{clock_part}
    \eea
    This profile contains the fingerprint of different scenarios, specified by the index $p$. $K$ is as defined above;
    $\omega$ is the frequency of the background oscillation induced by the Standard Clock in unit of the Hubble parameter $H_0$ ($H_0$ is evaluated at the time of sharp feature $t_0$); $k_r$ denotes the first resonant $K$-mode at $t_0$; $C$ is the amplitude;
    $\varphi$ is a constant phase, whose value depends on different correlation functions and models.
    For expanding scenarios, $K>k_r$; for contracting scenarios, $K<k_r$.
    The fingerprint resonant running refers to $\sin[\dots]$, of which the functional form of the argument is simply the inverse function of $a(t)$. By measuring this functional form, we know $a(t)$ directly, so this is the most important part of the clock signal.
    The envelop behavior specified by the parameter $\alpha$ can be model-dependent even within a scenario, but its overall scale-dependent trend can be used as a useful auxiliary evidence for a specific scenario.\footnote{Take the inflationary scenario for example. If some multi-field couplings in the model have certain strong scale-dependence, they can cause complicated scale-dependence in this envelop. But since the oscillation of the massive field quickly decays away for all inflationary models, the overall behavior of this envelop is a specific decaying behavior towards shorter scales.}  In the examples in this letter we will take $\alpha=-3/2+1/(2p)$ for power spectrum for reasons that we will explain.
\item[B.] {\em A relation between two scales.} Although the sharp feature signal (\ref{sin_part}) itself
    cannot be used to measure $a(t)$, its relative location to the clock signal is determined by $p$ and distinctive for different scenarios. As define above, let us denote $k_0$ as the starting $K$-location of sharp feature signal (\ref{sin_part}) and $k_r$ as the starting $K$-location of the clock signal (\ref{clock_part}). We have the following relation,
    \bea
    \frac{k_r}{k_0} = \frac{|p|}{|1-p|} \omega ~.
    \label{k_relation}
    \eea
    Note that this relation relies on the knowledge of the value of $\omega$, $p$ and $k_r$ which would be determined by fitting (\ref{clock_part}) to data, so it is closely connected to Property A. Once the parameters in (\ref{clock_part}) are determined, the relation (\ref{k_relation}) can be used to predict the location of the sharp feature $k_0$.
\end{enumerate}

\begin{figure}[t]
  \centering
  \includegraphics[width=1\textwidth]{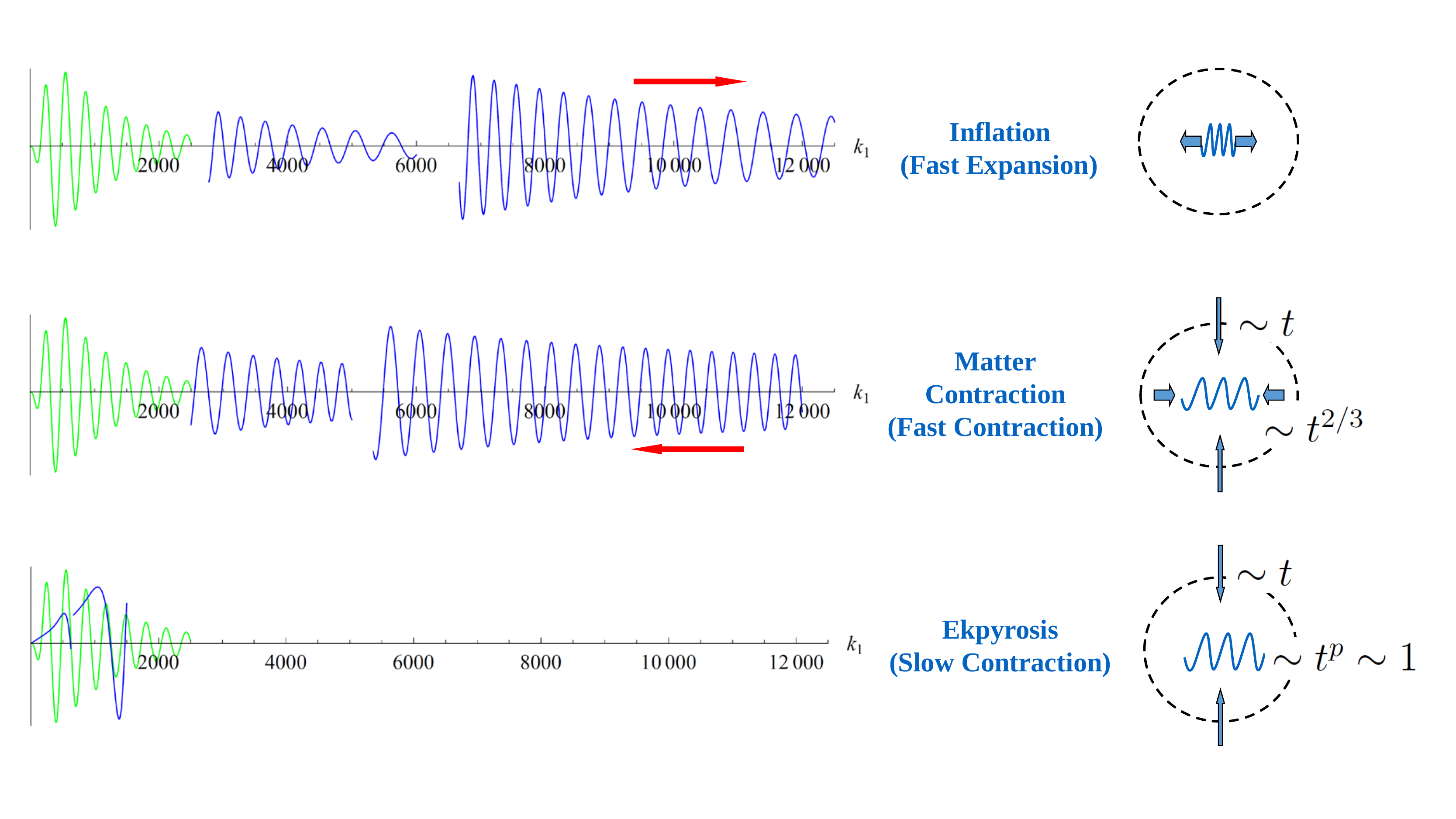}
\caption{A qualitative illustration of the full Standard Clock signals for different scenarios. Green/light lines represent the sharp feature sinusoidal running signals; blue/dark lines represent the clock signal with two different masses. Here the envelop behaviors are somewhat different from those in Ref.~\cite{Chen:2011tu,Chen:2012ja}, because we plot the direct coupling case here. Both key Properties remain the same. Also see text for explanations.}
\label{Fig:paradigms}
\end{figure}

As we explain below, both key Properties are simple and direct consequences of the scale factor evolution $a(t)$. We also illustrate this in Fig.~\ref{Fig:paradigms}.

We start with the Property A. We first note that the oscillating massive field sets a background frequency that is approximately constant.\footnote{In realistic models, the mass usually has some reasonable time dependence. This does not change the qualitative features we emphasize here.}
The amplitudes of the oscillations also have different characteristic time-dependent trends in different scenarios.

Let us first compare inflation with fast-contracting scenarios, such as the matter contraction \cite{Wands:1998yp,Finelli:2001sr}. The physical frequency of quantum fluctuations scales inversely proportional to the scale factor. When this frequency passes through the background frequency, the mode resonates with the background and generates a large signal \cite{Chen:2008wn}. For inflation the following is the evolutionary sequence for each mode: the physical frequency decreases, hits the resonance, and then the mode exits the horizon. So longer modes resonate first, the resonance starts from $K=k_r$ and runs towards larger $K$.
The ticks in the resonant running are created by the ticks of the Standard Clock.
Note that for inflation the relative distance between the ticks in the $K$-space increases as $K$ increases.
For fast-contracting scenario, the situation is the opposite. The physical frequency of modes increases before hitting the resonance. Modes are still exiting the horizon because the horizon contracts faster than the scale factor. So shorter modes resonate first. The resonance starts from $K=k_r$ and runs towards smaller $K$. Note that the relative distance between ticks increases as $K$ decreases, in contrast to the inflationary case.

Next let us compare inflation with slowly-varying scenarios, such as the Ekpyrosis \cite{Khoury:2001wf}.
For inflation, the horizon size is approximately constant, so all modes resonate sooner or later. We see many oscillations in the resonant signals.
For Ekpyrosis, the scale factor is evolving very slowly, so different modes hit the resonance in a very slow one-by-one fashion. However, the horizon size contracts fast. Once the modes exit the horizon, it can no longer resonate. So very few modes resonate, and that is why we see very few oscillations in the clock signal.

The Property B is also a direct consequence of $a(t)$. The ratio between the mode $k_0$ (which crosses the horizon at the time of sharp feature $t_0$) and the mode $k_r$ (which is deep inside the horizon and is the first mode to resonate at $t_0$) is determined by the ratio between the physical clock frequency $\Omega$, which we define as $\Omega\equiv\omega H_0$, and the horizon-mass-scale. The horizon-mass-scale ($(p-1)/t_0$) and $H_0=|p|/|t_0|$ are determined by the index $p$. So this leads to the relation (\ref{k_relation}) which depends on $p$ and $\omega$ degenerately.

These two properties are very robust and do not depend on many of the model details. For example, the inflation models can be either small field or large field; the massive fields can be excited by different kinds of sharp features; the coupling between the Standard Clock and the density perturbation source field can be gravitational or direct.

From (\ref{clock_part}), we can also understand the Requirement 2 for the model-building. Using (\ref{k_relation}), at $K=k_r$, the period of (\ref{clock_part}) is given by $\Delta K=2\pi k_0$. If the massive field starts to oscillate at or before the beginning of the observable epoch, in terms of the CMB multipole, $\ell_0 \lesssim 1$, then the period of (\ref{clock_part}) falls within $\Delta \ell \lesssim 2 \pi$, which is at or below the CMB resolution threshold. For inflation, $\Delta\ell$ increases for $K>k_r$ but the amplitude decays.

\section{A Standard Clock candidate in CMB}
\label{Sec:Candidate}

\begin{figure}[t]
  \centering
  \includegraphics[width=0.9\textwidth]{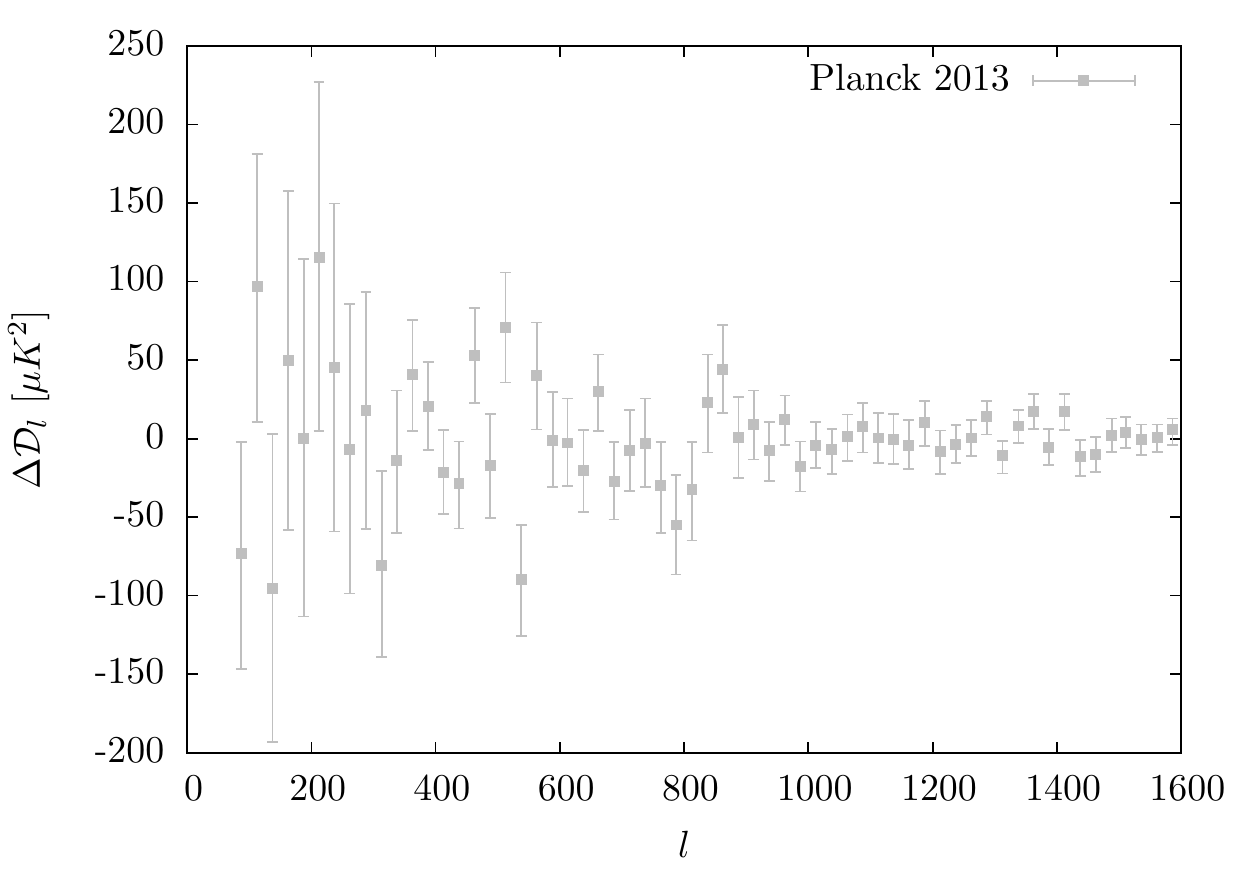}
\caption{Residuals of Planck 2013 binned temperature power spectrum from $\ell=100$ to $\ell=1600$, taken from \cite{Ade:2013zuv}.}
\label{Fig:Planckdata}
\end{figure}

Standard Clock signals show up as fine structures in the primordial perturbations and has better chance to be detected in high resolution maps measuring high multipoles, such as the Planck data.
Motivated by the theoretical predictions, we first search for the presence of the most important clock signal (\ref{clock_part}) in the Planck data. A detailed analysis will be presented elsewhere \cite{ToAppear}.
Here in this letter, we focus on an interesting candidate with a relatively low frequency, which is visible in the binned Planck data (Fig.~\ref{Fig:Planckdata}).

\begin{figure}[t]
  \centering
  \includegraphics[width=0.9\textwidth]{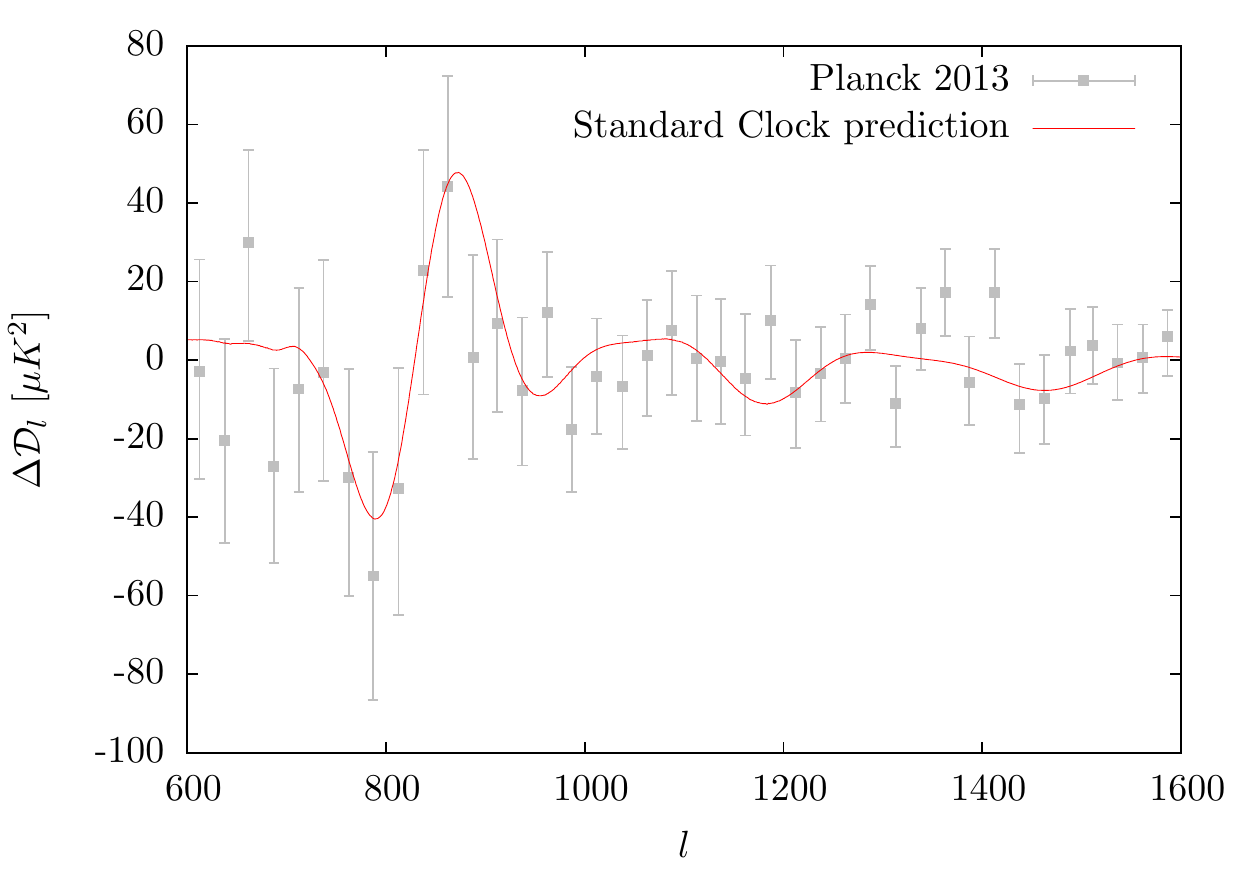}
\caption{A fit to a Standard Clock candidate using the clock signal profile (\ref{clock_part}) alone. The parameters are $\omega=30$, $C=0.075$, $k_r=0.1015~{\rm Mpc}^{-1}$, $p=100$, $\varphi=-0.263\pi$.}
\label{Fig:SC_fingerprint}
\end{figure}

We investigate if the oscillatory feature around $\ell\sim 800$ can be a viable candidate. Using the clock signal profile (\ref{clock_part}), we perform a Markov-Chain-Monte-Carlo analyses using Planck unbinned data varying all five model parameters in the neighborhood of this region; the ones with $p>1$ are preferred than the other values, and can improve the fit by $2\Delta \ln \CL \approx -9$ comparing to $\Lambda$CDM \cite{ToAppear}. In Fig.~\ref{Fig:SC_fingerprint}, we plot an example in comparison with the Planck residuals.\footnote{Note that all the figures illustrated in this paper are close to but not exactly the same as the rigorous best-fit examples. We leave the full data analyses, using MCMC search and vary all parameters including both the $\Lambda$CDM and Standard Clock parameters, to a forthcoming publication \cite{ToAppear}. The best-fit examples found by fine-tuning the parameters in the full analyses can increase the statistical significance slightly compare to the ones here. For example \cite{ToAppear}, one can get $2\Delta \ln \CL =9792.2-9802.8\approx -10.6$ comparing to $\Lambda$CDM by choosing $\omega=29.6$, $C=0.058$, $k_r=0.1075~{\rm Mpc}^{-1}$, $p=105$, $\varphi=0.684\pi$ (using the Planck and WMAP polarization data).

Also note that, although there are five parameters in the clock signal, if we were only interested in the inflation scenario, the number of effective parameters are four, because all $p\gg 1$ give the same result in practice, hence $p$ is no longer a parameter.}
The improvement is statistically insignificant so far. Nonetheless, we have several encouraging observations from this result.
\begin{itemize}
\item The oscillation in the data seems to decay very quickly as $\ell$ increases. But this behavior is well recovered by (\ref{clock_part}) through the transfer function, although the decay speed in the primordial profile (\ref{clock_part}) naively does not appear to be that fast.
\item The inflation case with $p\gg 1$ in (\ref{clock_part}) has a very specific running pattern which seems to be favored on the posteriors.
    The statistical improvement is mostly due to the fit to the wiggle at $\ell \sim 800$, and also due to the fact that the prediction in the region $\ell > 800$ does not contract with the data.
    If we were to take a sinusoidal wave with constant amplitude as a naive extension to the feature at $\ell\sim 800$, the fit would be worse in terms of both the amplitude and running pattern.
\item As we emphasized, the Standard Clock consists of two types of signals. Plugging the best-fit parameters for this clock signal in the relation (\ref{k_relation}) with $p\gg 1$, we can predict the $k_1$-location of the sharp feature signal $k_1=k_0/2$. This turns out to be at $\ell_0\sim 23$. As we know since the WMAP data \cite{Peiris:2003ff}, at $\ell\sim 20-30$ there is a well-known sharp feature candidate. We knew that it may be fit by a feature with sinusoidal running \cite{Adams:2001vc}, although this was never conclusive due to the marginal statistical significance.
\end{itemize}

These observations suggest that the two well-separated features in the CMB angular power spectrum may have the common origin associated with the Standard Clock effect.
Encouraged by these observations, we proceed to construct an explicit Standard Clock model and work out the full theoretical prediction on the power spectrum, as a completion of the existing results summarized in Sec.~\ref{Sec:StandardClocks}. In particular, we would like to see the complete prediction in the region between $k_0$ and $k_r$, and use it to jointly fit both features. We will focus on the exponential inflationary case below. Full predictions for other scenarios are left for future works.

\section{A full model of Standard Clock}
\label{Sec:FullModel}

Although sharp feature plays an important role in the Standard Clock models, we find its model-building identity may be quite flexible without affecting many important properties, as long as it excites massive fields. In fact such processes may be present in many existing models in the literature, even though their roles as Standard Clocks were not realized.

The case most studied is the sharp bending trajectory \cite{Chen:2011zf,Chen:2011tu,Gao:2013ota,Noumi:2013cfa}. In this case, it is found that, if we only consider the gravitational coupling between the Standard clock field and inflaton for the clock signal \cite{Chen:2011zf}, the leading resonance contributions from the background oscillation and massive field quantum fluctuation cancel each other, leaving a signal with a small amplitude \cite{Gao:2013ota,Noumi:2013cfa}. However the leading resonance in bispectrum \cite{Chen:2011zf} does not get cancelled and remain large \cite{Noumi:2013cfa}. For this case, one has to first look into the non-Gaussianity in data. As mentioned in Ref.~\cite{Chen:2011zf}, while the gravitational coupling is the minimal case, direct couplings between the Standard Clock and inflaton are of course possible. The direct coupling case is simpler to compute but much more model-dependent. Nonetheless most importantly, it should be clear that both Property A and B of the Standard Clock do not depend on these details. Examples of the direct coupling case are studied in Ref.~\cite{Saito:2012pd,Kobayashi:2012kc}.

For simplicity, in this letter we consider another type of sharp feature, namely the tachyonic falling, and we consider the direct coupling case. In this type of models, the inflaton is slow-rolling on its potential all the time; and the Standard Clock field starts to oscillate simply because it falls tachyonically into a potential dip and settles down. To satisfy the model building Requirement 2, we have to put the Standard Clock field on the plateau of the potential dip and let it (slow-)roll for at least a few efolds. This period corresponds to the density perturbation regime $\ell\lesssim 30$. The Lagrangian of the full model is
\bea
\CL =
-\frac{1}{2} (\tR + \sigma)^2 g^{\mu\nu} \partial_\mu \theta \partial_\nu \theta - V_{\rm sr}(\theta)
-\frac{1}{2} g^{\mu\nu} \partial_\mu \sigma \partial_\nu \sigma - V_\sigma(\sigma) ~,
\eea
where the potential
\bea
V_\sigma = V_0 \left[ 1- \exp(-\sigma^2/\sigma_f^2) \right]
\eea
models the potential dip for the clock field $\sigma$. $V_{\rm sr}$ can be any slow-roll potential for the inflaton field $\theta$. For example, for small field inflation we can approximate its total magnitude by a constant and the slope by a linear term, $V_{\rm sr} = V_1 - \beta \theta $; for large field, it can be simply the chaotic inflation potential $V_{\rm sr} = m^2\tR^2\theta^2/2$ \cite{Linde:1983gd}.
$\tR$ is approximately the final stable location of the massive field up to a small displacement due to a centrifugal force.
This Lagrangian is very similar to the quasi-single-field inflation model \cite{Chen:2009we,Chen:2009zp} with a large mass term \cite{Chen:2012ge,Pi:2012gf}, except that we now include an earlier phase describing how the $\sigma$ field slowly rolls and then drops and settles down in the potential dip. The curving of the trajectory introduces a direct coupling between the two fields.

Such a model is also natural from the model building point of view. It is known that some finetuning is required to have 60-efolds of inflation, hence many scalar fields fall down to their potential minima after a few efolds of rolling.

\begin{figure}[H]
  \centering
  \includegraphics[width=0.85\textwidth]{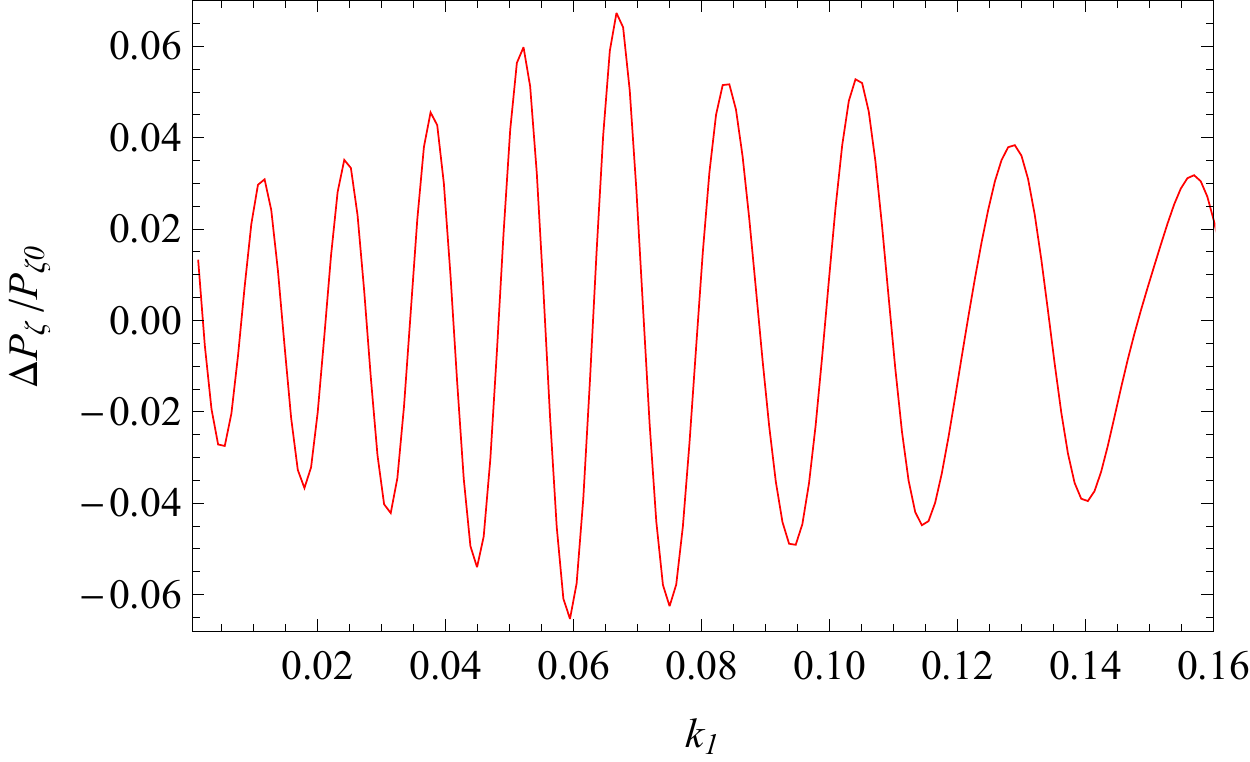}
  \includegraphics[width=0.90\textwidth]{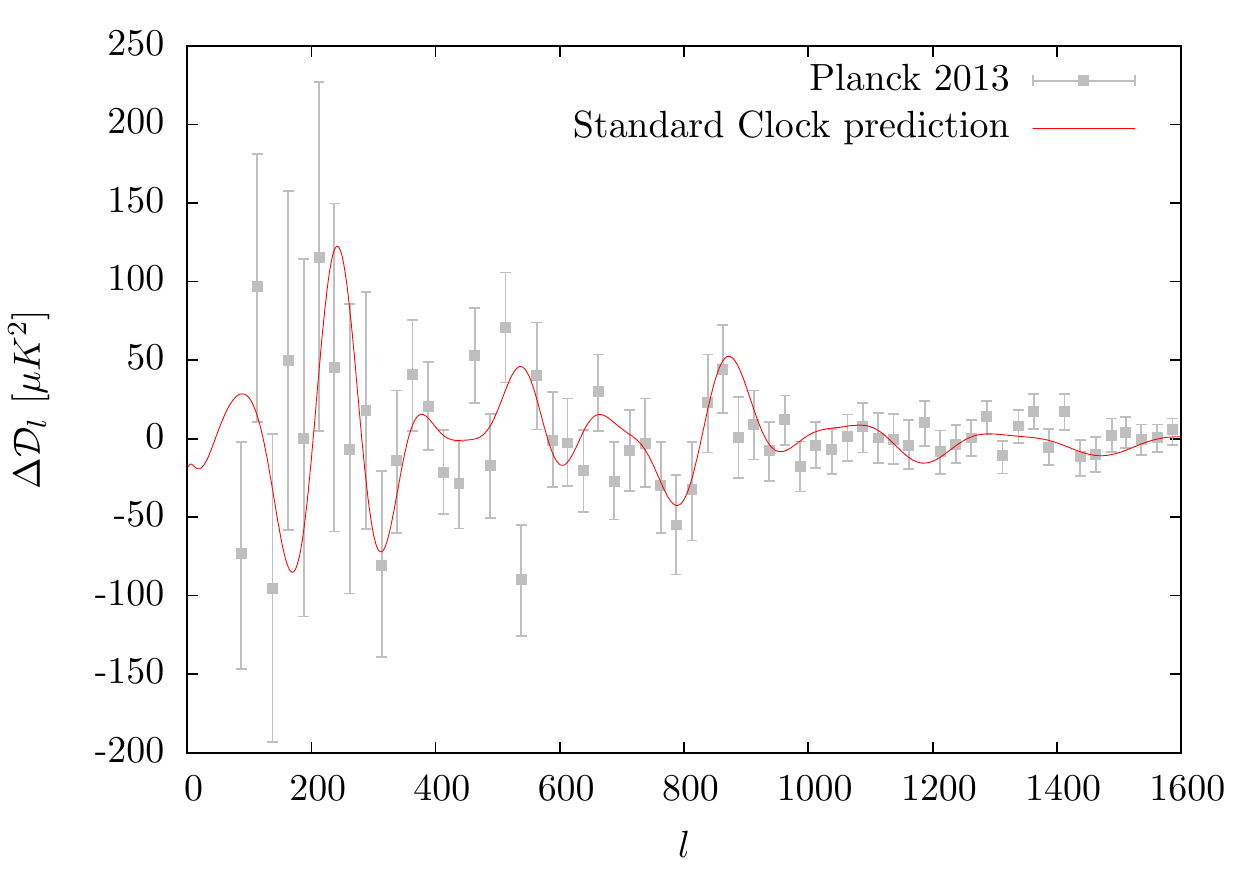}
\caption{A Standard Clock full numerical prediction using a small field example ({\it top}) and its comparison with the Planck residuals using the template (\ref{template}) ($k_r=0.106~{\rm Mpc}^{-1}$, $A=0.07$) and CAMB \cite{Lewis:1999bs} ({\it bottom}).}
\label{Fig:FullModel_smallfield}
\end{figure}

The background evolution can be solved numerically, and we denote it as $\theta_0(t)$ and $\sigma_0(t)$. Similar to the bending trajectory case, the feature effect on the inflaton fluctuations $\delta\theta$ (including both sharp feature and resonance effect) includes the contribution from the background oscillation and the contribution from the quantum fluctuations of the clock field $\delta\sigma$. The coupling of the latter is parameterized by $\dot\theta_0/\tR$. In the parameter space where this coupling is small, the dominant contribution simply comes from the background oscillation,
\begin{align}
\ddot{\delta\theta} +
\left[ 3H + \frac{2\dot\sigma_0}{\tR + \sigma_0} \right]
\dot{\delta\theta}
+
\left[ \frac{k^2}{a^2} + \frac{V''_{\rm sr}}{(\tR + \sigma_0)^2} \right]
\delta\theta =0 ~.
\label{bkgd_approx}
\end{align}
Setting Bunch-Davies vacua for both fields, we can solve the equation numerically \cite{ToAppear}. An example is presented in Fig.~\ref{Fig:FullModel_smallfield}. In this example we use the small field inflation model, with parameters $\mpl=1,~ V_1=1.66\times 10^{-13},~ V_0 = 5.33\times 10^{-15},~ \sigma_f=0.0164,~ \tR=2.05,~ \beta=1.38\times 10^{-16}$. We always choose parameters so that $\omega=30$ to reproduce the partial result in Fig.~\ref{Fig:SC_fingerprint}.

Notice, for the clock signal (\ref{clock_part}), there are some differences between the direct coupling case here and the gravitational coupling case \cite{Chen:2011zf,Chen:2011tu}. In the gravitational coupling case, the inflaton couples to the background slow-roll parameters, whose oscillating frequency is determined by the oscillating frequency of the energy density of the Standard Clock field. At the leading order, the energy is conserved, so $\omega=2m/H_0$ is given by the next order term. The value $\alpha = -3+5/(2p) \approx -3$ reflects the rate the density of the massive field gets diluted in the inflationary background. Here in the direct coupling case, the massive field $\sigma$ directly couples to the inflaton, instead of through the energy-momentum tensor. As a consequence, $\omega=m/H_0$ and $\alpha =-3/2 + 1/(2p) \approx -3/2$. Again none of these affects the Property A and B.

To compare with data, we use the following template to represent the numerical result. Differences between the numerical result (the upper panel in Fig.~\ref{Fig:FullModel_smallfield}) and the template are negligible on the scales of interest.
\bea
\frac{\Delta P_\zeta}{P_{\zeta0}} =
\left\{
\begin{array}{ll}
\high{
A \left[ 7\times 10^{-4} \left( \frac{2k}{k_0} \right)^2 + 0.5 \right]
\cos \left[ \frac{2k}{k_0} + 0.55\pi \right] ~,
}
&
k<k_a ~,
\\
\high{
\frac{14}{13} A \left( \frac{2k}{k_r} \right)^{-3/2}
\sin \left[ \omega \ln \frac{2k}{k_r} + 0.75\pi \right] ~,
}
&
k_b >k\ge k_a ~,
\\
\high{
\frac{19}{13} A \left( \frac{2k}{k_r} \right)^{-3/2}
\sin \left[ \omega \ln \frac{2k}{k_r} + 0.75\pi \right] ~,
}
&
k\ge k_b ~,
\end{array}
\right.
\label{template}
\eea
where
\bea
k_0 = \frac{k_r}{1.05\omega} ~, ~~~
k_a = \frac{67}{140}k_r ~, ~~~
k_b = \frac{24}{35}k_r ~, ~~~
\omega=30 ~.
\eea
As we can see, the transition from the sharp feature sinusoidal running to the clock resonant running illustrated in Fig.~\ref{Fig:paradigms}, and the Property A and B -- all with the inflationary characters -- are manifest in this template.

Unlike (\ref{clock_part}), for the full result, we have not figured out a way to write down the analytical template with varying $\omega$. So to compare with data, we simply choose and fix one of the best-fit values from Sec.~\ref{Sec:Candidate}, $\omega=30$. There are only two free parameters in this template, the scale $k_r$ and the amplitude $A$.
Visually a fit is demonstrated in Fig.~\ref{Fig:FullModel_smallfield}.
It is worth to emphasize that, the total number and the value of parameters for the full prediction are the same as those of the clock signal in Fig.~\ref{Fig:SC_fingerprint}. No new parameters are introduced while the amount of the prediction is increased significantly. Therefore, it is encouraging that, between the two scales $\ell_0\sim 30$ and $\ell_r\sim 700$, the full theoretical prediction not only does not contradict with the data, but also shows some fitting behavior to the otherwise noise-looking residual data.\footnote{However the statistical improvement in addition to Fig.~\ref{Fig:SC_fingerprint} gained in the majority portion of the range $\ell\sim 100-700$ is cancelled (back to $-2 \Delta\ln \CL \approx -7.8$ \cite{ToAppear}) by data from a specific scale, represented by the single binned data point at $\ell = 538$ in Fig.~\ref{Fig:FullModel_smallfield}. This data point alone increases $-2 \Delta\ln \CL$ by $\sim 8$. The future determination of this data point will be important for this candidate.}

\begin{figure}[H]
  \centering
  \includegraphics[width=0.90\textwidth]{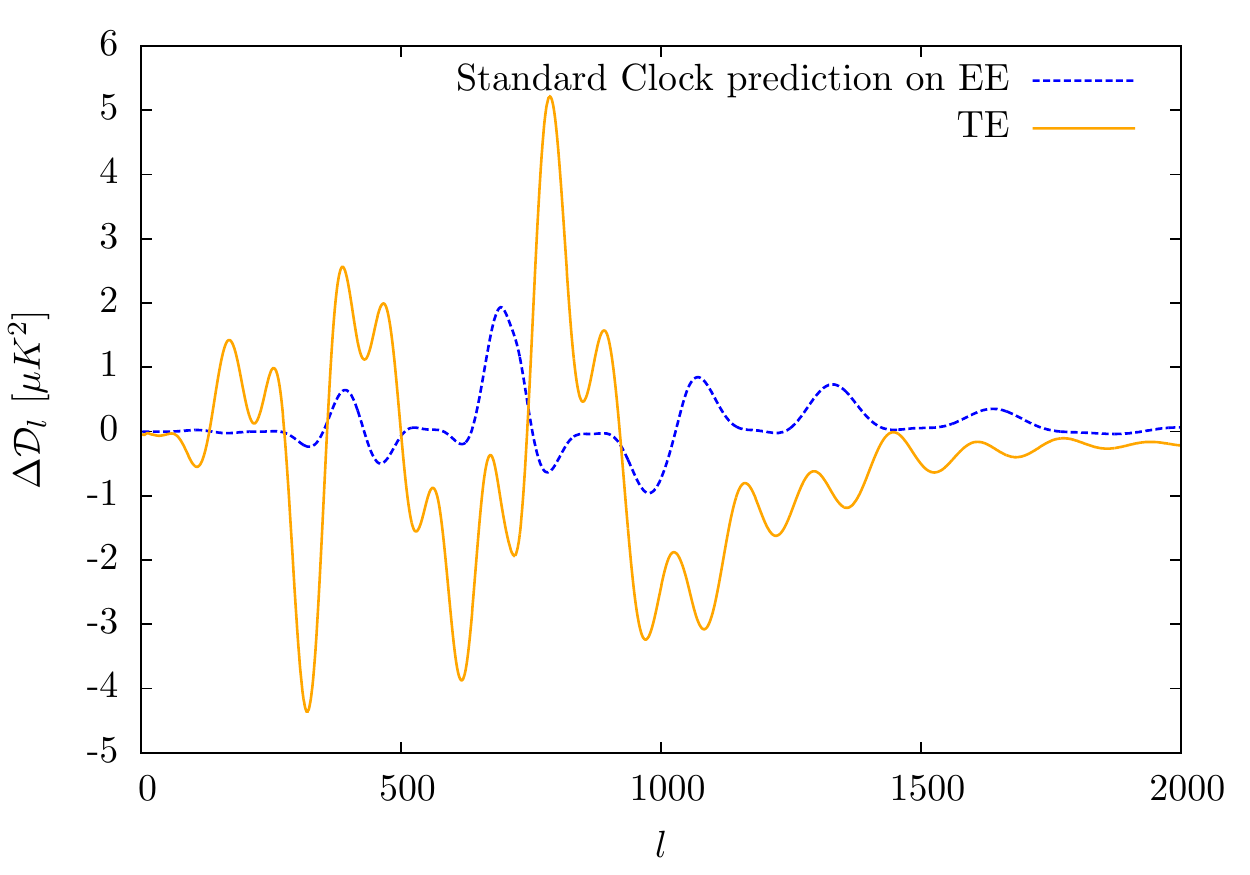}
\caption{Predictions on EE and TE residuals correlated with Fig.~\ref{Fig:FullModel_smallfield}.}
\label{Fig:smallfield_TE}
\end{figure}

The full theoretical result also predicts the highly correlated signals in the EE and ET polarization residuals that can be tested by future Planck data, see Fig.~\ref{Fig:smallfield_TE}.

We have also tested several large field inflation examples. As we increase the value of the slow-roll parameter $\epsilon$ towards large field models, the most significant change is the increasing amplitude of the envelop of the sharp feature signal. This modifies the relative amplitude sizes between the sharp feature signal and the clock signal, and can significantly affect the data fitting. We leave the detailed study to Ref.~\cite{ToAppear}.

\section{Conclusion}

Following the theoretical proposal of Standard Clock, we have started to compare its predictions with the Planck data. We have constructed and worked out a Standard Clock model with full predictions on the power spectrum. We have presented an interesting candidate in the Planck data, characteristic of the Standard Clock in the inflationary scenario.
Although this candidate is not yet statistically significant, we use this to motivate detailed theoretical model-building, and use it as an example to illustrate how Standard Clock appears in CMB and how they can be further tested by future data.
Such a Standard Clock candidate is an extensive feature covering nearly the entire range of scales probed by Planck. It would potentially add at least four more parameters to the $\Lambda$CDM Standard Model. It has highly correlated predictions in the polarization map, non-Gaussianities and other large scale structure maps over the same wide range of scales, and so can be tested with further analyses and future data. The Standard Clock signal contains important information on the early universe, and in particular can be used to directly measure the time-dependence of the scale factor of the primordial universe. If any of these candidates is verified, such a signal would provide an independent and direct evidence for the inflationary paradigm.

\medskip
\section*{Acknowledgments}

We would like to thank Christophe Ringeval and Yi Wang for valuable help on programming, data search and collaboration \cite{ToAppear}.
We thank Hassan Firouzjahi, Xian Gao, David Langlois, Shuntaro Mizuno and Misao Sasaki for helpful discussions. Part of this work was presented in the Mini-Workshop on Gravitation and Cosmology (Feb 6th, 2014) in Yukawa Institute of Theoretical Physics in Kyoto University \cite{YITPworkshop}. XC would like to thank the organizers for the hospitality.


\end{spacing}

\end{document}